\documentclass[galaxies,article,accept,moreauthors,LaTeX and dvi2pdf]{Definitions/mdpi} 
\firstpage{1} 
\pubvolume{xx}
\issuenum{1}
\articlenumber{5}
\pubyear{2020}
\copyrightyear{2020}
\history{Received: 30 June 2020; Accepted: 04 August 2020; Published: date}

\usepackage{subcaption}
\usepackage[countmax]{subfloat}
\Title{Temporal and Spectral Variability of OJ 287 before the April--June 2020 Outburst}

\Author{Nibedita Kalita 
 $^{1,}$* \orcidA{}, Alok C. Gupta $^{2}$ \orcidB{}, and Minfeng Gu $^{1}$} 

\AuthorNames{Nibedita Kalita}
\address{%
$^{1}$ \quad Key Laboratory for Research in Galaxies and Cosmology, Shanghai Astronomical Observatory, Chinese~Academy of Sciences, 80 Nandan Road, Shanghai 200030, China; gumf@shao.ac.cn\\
$^{2}$ \quad Aryabhatta Research Institute of Observational Sciences (ARIES), Nainital 263001, India; acgupta30@gmail.com}

\corres{Correspondence: nibeditaklt1@gmail.com}

\abstract{We present the results of a temporal and spectral study of the BL Lacertae object OJ 287 in optical, UV, and X-ray bands with observations performed by {\it Swift} satellite during September 2019--March 2020. In~this period, the source showed moderate variability characterized by variability amplitude of $\sim$22--31\% in all the wavelengths on a short timescale, except the hard X-ray band which was variable by only $\sim$8\%. We~observed that the X-ray flux of the source was significantly dominated by the soft photons below $2$ keV. Soft~lags of $\sim$45~days were detected between the optical/UV and soft X-ray emissions, while there is no correlation between the hard X-rays and the lower energy bands indicating the presence of two emission components or electron populations. Although two components contribute to the X-ray emission, most of the 0.3--10~keV spectra were well fitted with an absorbed power-law model which outlines the dominance of synchrotron over inverse Compton (IC) mechanism. The X-ray spectra follow a weak ``softer when brighter''~trend.}


\begin{document}

\section{Introduction}\label{sec:intro}
\noindent

The BL Lacertae (BL Lac) object OJ 287 (0851 + 203; z = 0.306) is one of the brightest and most frequently observed blazar. OJ 287 shares the common BL Lac properties of blazar class which include significant flux variation in the complete electromagnetic (EM) spectrum on diverse timescales, variable radio to optical polarization, superluminal motion of the compact radio core, and predominantly non-thermal emission (e.g., \cite{2018MNRAS.473.1145K, 2019AJ....157...95G, 2019ApJ...882...88V}  and references therein). Blazars emit relativistic charged particles in the form of collimated jets pointing towards the observer's line of sight \citep{1995PASP..107..803U}, and their radiation covers a wide range of frequencies giving us an excellent opportunity to study their broadband spectral energy distributions (SEDs) which can help us to understand the hidden physical scenario in their local environment. The relativistically boosted SEDs of blazars display two broad humps, where the low energy
hump peaks at radio to X-ray wavelengths and the high energy hump appears in the $\gamma$-rays ranging from MeV to TeV energies. The first SED hump is ascribed to synchrotron emission from a relativistic jet, while the second hump arises probably from IC scattering of lower energy photons off the synchrotron emitting relativistic charged particles (e.g., \cite{1995PASP..107..803U, 1998A&A...333..452K, 2010ApJ...718..279G} and references therein). 

OJ 287 has more than a century-long optical observations starting from the year 1890, which is the longest optical observations of any blazar. It was noticed from quasi-periodic analysis of those observations that the source showed a double-peaked optical outburst at every $\sim$12 years~\cite{1988ApJ...325..628S}, ascribed to a binary black hole (BH) model, where the secondary BH crosses the accretion disc of the primary twice in every 12 years to complete its orbit, which triggers two outbursts during each impact. In~the~past, predictions for periodic outbursts were made based on this theory and which were indeed detected (e.g., \cite{1996A&A...305L..17S,1996A&A...315L..13S, 2009ApJ...698..781V}). Studies found a peak separation of $\sim$1.2--2 years in the double-peaked outbursts. From these periodic events, the masses of the primary and the secondary supermassive black holes (SMBHs) are estimated to be (1.83 $\pm$ 0.01) $\times$~10$^{10}$ M$_{\odot}$ and (1.5 $\pm$ 0.1) $\times$ 10$^{8}$ M$_{\odot}$, respectively, and the spin of the primary SMBH to be 0.313 $\pm$ 0.01~\citep{2016ApJ...819L..37V}. Outburst in OJ 287 is predicted for 2015--2020 in which the first peak was observed in December 2015~\citep{2016ApJ...819L..37V, 2017MNRAS.465.4423G}, and the second peak of the outburst is yet to be detected. Recently, the source has undergone an outburst at optical, UV, X-ray, and radio wavelengths during April--June 2020 (ATel: 13677, 13702, 13755, 13785, 13658), which may be the second predicted outburst?

In OJ 287, there are several peculiar and interesting features observed on different occasions. The~broadband SED of the source extends up to very-high-energy (VHE) $\gamma$-rays (>100 GeV) and belong to the low synchrotron peaked (LSP) blazar category. However, in a recent study during a VHE activity period \citep{2017ICRC...35..650B}, the source showed HBL like components, where the SEDs were described by a two-zone leptonic model, being the second zone located at parsec scales, beyond the broad line region \citep{2018MNRAS.479.1672K}. The~source shows diverse spectral behavior in the X-ray energy band, covering extremely soft spectral profiles to a very flat one or a combination of both. It has been observed that the source goes to an extremely soft X-ray state usually around the period of optical outbursts \citep{2018ApJ...866...11D}. A signature~of optical-UV bumps in the multi-wavelength SEDs was observed during the December 2015 to May 2016 flaring activity of the source \citep{2018MNRAS.473.1145K}. Recently, Pal et al. \cite{2020ApJ...890...47P} reported a significant soft excess below 2 keV compared to its generally observed power-law spectrum in the X-ray band (0.3--10~keV) with {\it XMM-Newton} observations. The systematic study found that this excess can not be due to the extension of the synchrotron component to X-rays. Based on optical to X-ray modeling, they~found X-ray reprocessing or cool Comptonization as the best candidates but favored the reprocessing scenario on the basis of lagging of UV with respect to X-ray. Both the interpretations were consistent with the observed NIR-optical break, argued to be a standard accretion disk emission of a 10$^{10}$ M$_{\odot}$ primary~SMBH. 

\begin{figure}[H]
\centering
\includegraphics[scale=0.8]{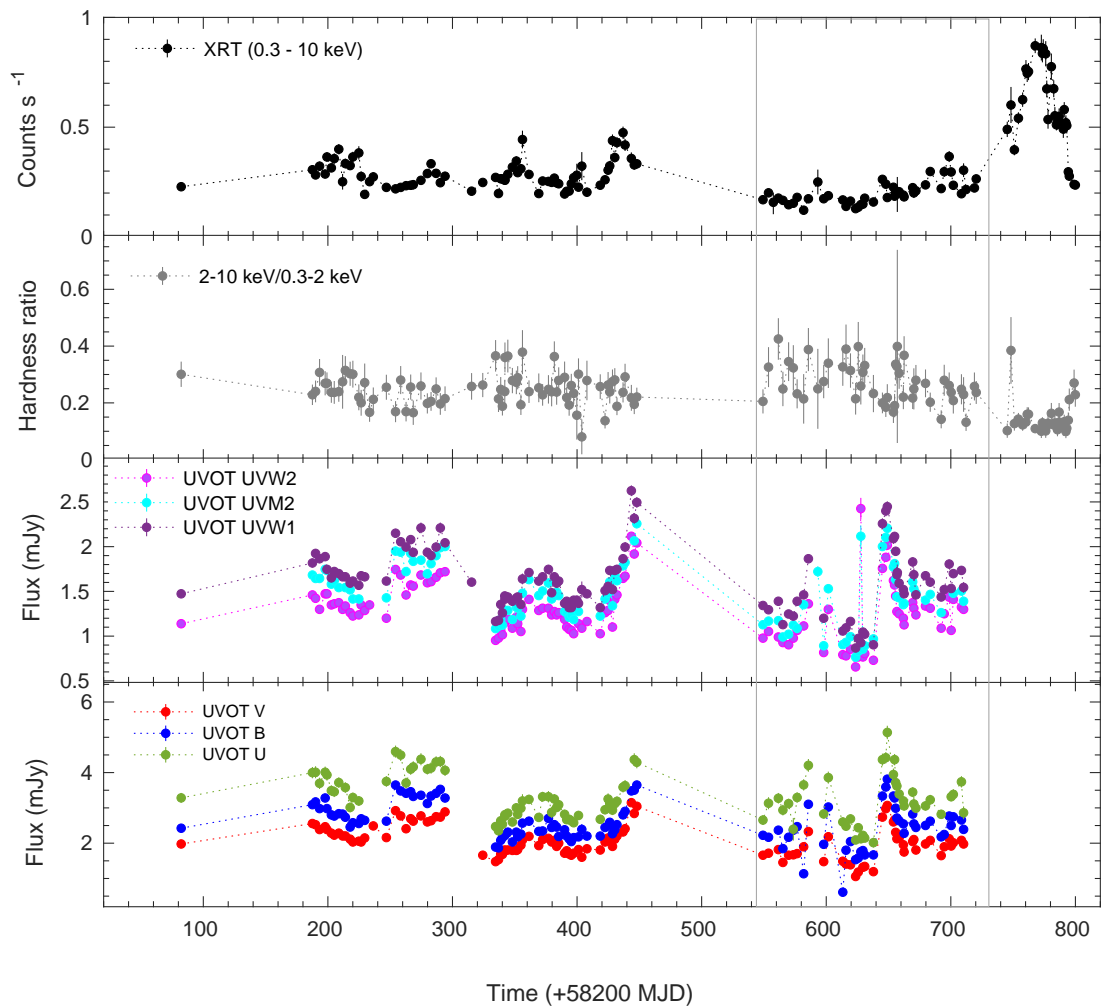}
\caption{Multi-wavelength light curves of OJ 287 from September 2018 to June 2020 observed by {\it Swift}---XRT and UVOT  telescopes. From top to bottom, the panels show XRT light curve in the energy range of 0.3--10 keV, X-ray hardness ratio, UV light curves in UVW2, UVM2 and UVW1 filters and the bottom panel shows the optical light curves in U, B, and V bands, respectively. All of the light curves were binned at intervals of observation ID. The pre-outburst area is indicated by vertical lines.} 
\label{fig1}
\end{figure}

In the present work, we focus on temporal and spectral evolution of the blazar in the pre-outburst period shown between the two vertical lines in Figure \ref{fig1}. The X-ray light curve of the blazar OJ 287 was generated using observations between June 2018 to June 2020 and is shown in Figure \ref{fig1}, along with the UV--optical light curves from June 2018 to March 2020.

The paper is organized as follows. Section \ref{sec:data} describes the data processing and analysis methods. In~Section \ref{sec:results}, we explain the 
results of a timing and spectral analysis, and Section \ref{sec:summary} provides the summary of this work.


\section{Swift Observations and Data Reduction}\label{sec:data}
\noindent

We collected {\it Swift} data from two on-board instruments: the X-ray Telescope (XRT; Burrows et~al.,~2005) and the UV and Optical Telescope (UVOT; \cite{2005SSRv..120...95R}) for the blazar OJ 287 between MJD 58282 to~58920.

All the XRT observations taken during the mentioned period were carried out in the most sensitive photon counting readout mode. We analyzed the pointed observations using the {\it XRTDAS} software package distributed within the {\it HEASOFT} package. Event files were cleaned and calibrated by applying standard filtering criteria with the {\it xrtpipeline} task and using the latest calibration files. In~order to extract the source and background spectrum, we used a circular source region of 47.2~{\it arcsec} and an annular source-free region, around twice the size of the source region as a background within the {\it xselect} interface as described in Kalita et al. \cite{2019ApJ...880...19K}. In some observations, the source count rate was high enough ($\geq$0.5 counts s$^{-1}$) to cause some photon pile-up in the inner pixels of the source region, which were corrected by removing those central pixels by carefully examining the PSF plot via ximage task. Ancillary response files for the spectral analysis were generated with the {\it xrtmkarf} task applying corrections for the PSF and damaged CCD pixels. The spectra were grouped to include at least 20~counts per energy bin in order to make it suitable for $\chi$$^{2}$ fitting. $\chi$$^{2}$ statistics  is  strictly  defined  for  Gaussian-distributed data and the particular grouping of the spectra is required to minimize the deviation of data from Gaussianity \citep{2009ApJ...693..822H}.

UVOT observations taken simultaneously with the XRT were available for all the filters (U, B, V, UVW1, UVM2, and UVW2) for most of the pointings. We derived the sky-corrected images from the {\it Swift} archive and then applied the task {\it UVOTIMSUM} to combine individual exposures if they exist for a single filter in an observation. The aperture photometry was performed with the {\it UVOTSOURCE} task with large scale sensitivity correction, using a source region of 5 {\it arcsec} and an  annular background region having inner and outer radii of 25 {\it arcsec} and 35 {\it arcsec} centered on the source. Before extracting the source flux, we checked for bad pixels on the UVOT detector using the small scale sensitivity file available online \footnote{\url{https://swift.gsfc.nasa.gov/analysis/uvot\_digest/sss\_check.html}}. The fluxes were dereddened using a value for E(B-V) of 0.028 $\pm$ 0.0008 with the recipe given by Roming et al. \cite{2009ApJ...690..163R}. 

Multi-wavelength light curves of the source are shown in Figure \ref{fig1}. The 0.3--10 keV XRT light curve is shown in the top panel of the figure, and the next panel shows the hardness ratio (HR), which is the ratio of hard (2--10 keV) to soft (0.3--2 keV) X-ray count rates, and the optical and UV light curves in different UVOT filter bands are displayed in the lower panels. All these light curves were constructed using time bins of observation IDs, where each bin contains around 80--170 counts and the typical exposure time of these observations lies in the range of 600--1200 s. The vertical solid lines dividing the plot denote the pre-outburst period, September 2019--March 2020, that we have studied in this work. During this epoch, the source was targeted 45 times by the {\it Swift} observatory.

\section{Results}\label{sec:results}
\subsection{Short-Term Variability}
\noindent

Blazars show flux variations in all the energy bands on diverse time-scales. Those fluctuations ranging from a few minutes to less than a day in the observer’s frame are often known as intra-day variability (IDV; \cite{1995ARA&A..33..163W}). Variability with time-scales of few days to weeks is commonly known as short-term variability (STV), while variability on time-scales ranging from months to years is called long-term variability (LTV; \cite{2004A&A...422..505G}).

In order to estimate the strength of variability, we used the fractional rms variability amplitude ($F_{var}$) method \citep{2002ApJ...568..610E, 2015MNRAS.451.1356K}, which gives the average variability amplitude of a time series data with respect to its mean flux. If a time-series has $N$ estimates flux $x_i$, with corresponding finite uncertainties $\sigma_{err,i}$ related to the measurement errors, then the excess variance is expressed as

\begin{equation}
\sigma_{XS}^2 = S^2 - \bar\sigma_{err}^2
\end{equation}
where $\bar{\sigma_{err}^2}$ is the mean square error, defined as
\begin{equation}
\bar\sigma_{err}^2 =\frac{1}{N} \sum\limits_{i=1}^N \sigma^2_{err,i} 
\end{equation}
and $S^2$ is the sample variance of the LC, given by
\begin{equation}
S^2 = \frac{1}{N-1} \sum\limits_{i=1}^N (x_i - \bar{x})^2 
\end{equation}
where $\bar{x}$ is mean of $x_i$. 

Here, $\sigma^2_{NXS} = {\sigma^2_{XS}} / {\bar{x}^2}$ is the normalized excess variance
and the fractional rms variability amplitude, $F_{var}$, which is the square root of $\sigma^2_{NXS}$, is given as
\begin{equation}
F_{var} = \sqrt{\frac{S^2 - \bar{\sigma}^2_{err}}{{\bar{x}^2}}} 
\end{equation}
The uncertainty on $F_{var}$ is given by Vaughan et al. \cite{2003MNRAS.345.1271V}
\begin{equation}
err(F_{var}) =  \sqrt{\left( \sqrt{\frac{1}{2N}}\frac{\bar\sigma_{err}^2}{\bar{x}^2 F_{var}} \right)  ^ 2+ \left(  \sqrt{\frac{\bar\sigma^2_{err}}{N}} \frac{1}{\bar{x}}\right) ^2 }
\end{equation}

\noindent

From this analysis, we found that the source was variable in all the bands on a moderate scale with $F_{var}$ values 24.48 $\pm$ 0.51\%, 26.94 $\pm$ 0.54\%, 22.05 $\pm$ 0.78\%, 30.54 $\pm$ 0.44\%, 28.09 $\pm$ 0.52\%, and \mbox{26.28 $\pm$ 0.50\%} in optical U, B, and V bands and ultraviolet UVW2, UVM2, and UVW1 filters, respectively. The corresponding excess variances for these six bands are 0.22, 0.41, 0.49,  0.14, 0.14, and~0.16, respectively. In X-rays, the variability strength is 22.64 $\pm$ 1.82\% ($\sigma_{XS}^2 =$ 2.5 $\times$ 10$^{-3}$) in the total energy range, with soft X-rays being significantly variable (\mbox{$F_{var}$ = 25.06 $\pm$ 1.24\% $\&$ $\sigma_{XS}^2 =$ 1.92 $\times$ 10$^{-3}$}) as compared to that of hard X-rays ($F_{var}$= 8.28 $\pm$ 3.09\% $\&$ $\sigma_{XS}^2 =$ 5.74 $\times$ 10$^{-5} $ ). We noticed that, from optical to soft X-rays, the source shows more or less similar variability strength. A similar kind of result was found for the IBL, ON 231 in our previous study \citep{2019ApJ...880...19K} and also, for other IBLs (e.g., BL Lacs, S5 0716 + 714, 3C 66A and 4C + 21.35; \cite{2016MNRAS.458...56W, 2018MNRAS.480..407K}). This type of behavior in LBL/IBL sources is due to the presence of two different emission components; the synchrotron and IC components, in their broad-band spectra. Most of the time (usually in the bright phase of the object), these two components; the high energy tail of the first component and the low energy face of the second component meet somewhere in the high energy X-ray band, resulting in an upward turn in the spectra. In these sources, the flux variability in the synchrotron domain is more pronounced as compared to that in the IC domain. As we found in our study, the contribution from different emission components in the X-ray spectra of OJ 287 was reported in some previous studies as well (e.g., \cite{2008A&A...489.1047M,  2018MNRAS.479.1672K, 2018MNRAS.480..407K} and references therein). 

\subsection{Variability Correlation}
\noindent


To check the correlation between different energy bands, we first did a least-squares fitting between different wavebands and found that all the inter and intra-optical/UV bands are well correlated to each other having a Pearson coefficient, $R >$ 0.91 with a very low probability ($P$; close to zero) of observing the null hypothesis. Visually, a linear relationship between soft and optical-UV bands was not very clear, however, we got $R$ values ranging from 0.55 to 0.58 with probabilities in the order of $10^{-4}$. In this study, we did not see any relation between soft and hard X-rays ($R = 0.23$ and $P = 0.12$).

In order to achieve a more clear view, we applied the Discrete Correlation Function (DCF; \cite{1988ApJ...333..646E}). DCF is one of the best methods to investigate a correlation between two unevenly sampled time-series data. The main advantage of this method is its capacity to handle the data without interpolating it, resulting in more accurate outputs with proper error estimation. A detailed description of the method we used in this work can be found in Kalita et al. \cite{2019ApJ...880...19K}.

\begin{figure}[H]
\centering
\includegraphics[scale=0.5]{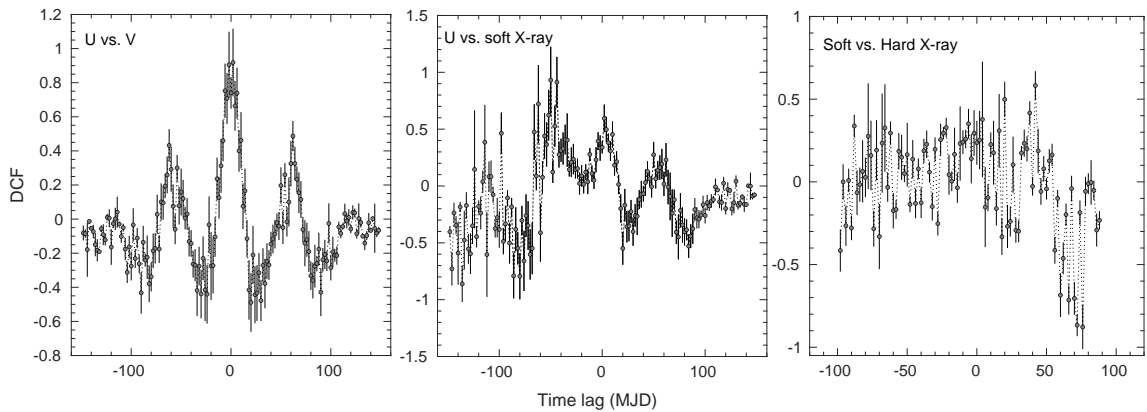}
\caption{Correlation between different energy bands using the DCF method. First, middle, and last panels represent DCF between optical U versus optical V bands, optical U versus UV UVW1 bands, and soft X-ray (0.3--2 keV) versus hard X-ray (2--10 keV) bands, respectively.} 
\label{fig2}
\end{figure}

We used this method to estimate any possible correlations between the optical, UV, and X-ray bands and also between the intra-bands. To do so, we measured DCFs between the soft X-ray and all the other energy bands and also between the optical and UV bands. We used a binning of 2 which is equivalent to a bin size of around 10 days. We found that all the optical and UV bands are nicely correlated to each other having DCF curve similar to that shown in the first panel of Figure \ref{fig2}, with~DCF peaks at $\approx$0.80--0.90 i.e., with 80--90\% degree of correlation. In this case, the cross-correlation is very nearly symmetrical around zero lag, which is almost identical to an auto-correlation. In~such~cases, except for a narrow region around zero lag, correlated fluctuations in the two time-series are statistically indistinguishable \citep{1988ApJ...333..646E}. We detected a soft lag of 18 MJDs between U and V bands; however, lags were less than the DCF bin size in other optical vs. UV bands. A negative lag of 45--48 MJDs between soft X-rays and all the optical/UV bands were found with DCF$_{peak}$s at around 0.70. Although~we got consistent results in this case, the peak values are not large enough to consider these as strong correlations (middle panel, Figure~\ref{fig2}). In X-rays, no correlation was observed between the soft and hard bands as there is no peak in the DCF curve shown in the last panel of Figure~\ref{fig2}. The peak values and time lags were estimated by fitting a Gaussian model of the following form to the DCF measurements: 

\begin{equation}
DCF(\tau)=a \times {\rm exp}\Bigl[\frac{-(\tau - m)^{2}}{2 \sigma^{2}}\Bigr] 
\end{equation}

Here, $m$, $a$, and $\sigma$ represent the time lag at which DCF peaks, the peak value of the DCF, and the width of the Gaussian function, respectively. 

From this analysis, we found that the optical soft X-ray emissions have correlated variability with soft lags if there is any. The exception is the hard X-ray emissions, which are totally independent of the rest. These results explicitly indicate that the emission up to soft X-rays and beyond this energy range are governed by two electron populations and mechanisms. The most probable scenario is that the lower energy photons were emitted by the synchrotron electrons in the jet, while the hard X-ray photons were emitted by IC emission. Usually, a weak correlation is present if the seed photons for IC scattering are coming from the jet itself. Thus, in this case, those photons were possibly originated from an external source.

\subsection{Spectral Variability}
\noindent

We performed the X-ray spectral analysis in the energy range of 0.3--10 keV using {\it XSPEC} 12.10.0c~\citep{1996ASPC..101...17A}, and used the $\chi^{2}$ statistic to determine the goodness of fits to spectra. In this analysis, error for each parameter was estimated from 2$\sigma$ (90\%) confidence ranges. To fit the spectra, we used either an absorbed power-law or a log-parabola model depending on the {\it ftest} probability, which compares the $\chi^{2}$ statistics for the two models and gives F-statistic and its probability. A lower value of probability recommends a higher likelihood of the latter model being a true representative of the data. For all the spectra, we found a neutral hydrogen column density ($N_{H}$) either below or equivalent to the Galactic value of $2.38 \times 10^{20}$ cm$^{-2}$. Thus, we freeze this parameter to the Galactic value for all the spectra. Finally, the task {\it cflux} was used to estimate the source flux from the best fit model. Due to inadequate data in some observations, we performed spectral fitting on 33 spectra out of 45 and, among those, only~in one instance was Log-parabola preferred over an absorbed power-law model.

\begin{figure}[H]
\centering
\includegraphics[scale=0.65]{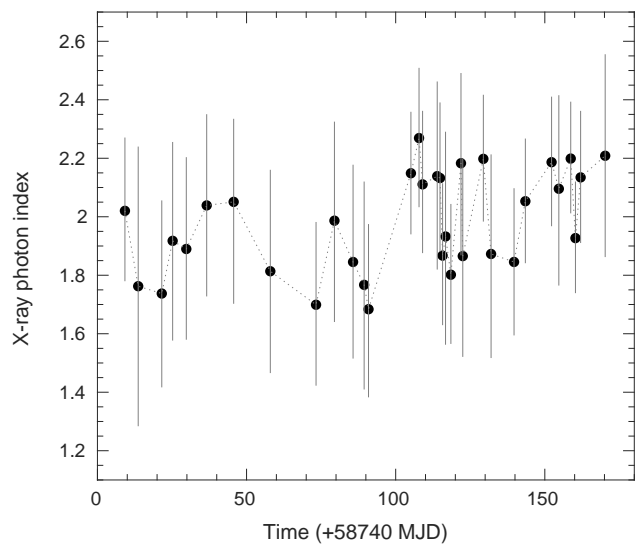}
\caption{Spectral evolution of OJ 287 before the April--June 2020 outburst. Photon indices are estimated from the 0.3--10 keV spectra by fitting an absorbed power-law or a log-parabola model.} 
\label{fig3}
\end{figure}

\begin{figure}[H]
\centering
\includegraphics[scale=0.65]{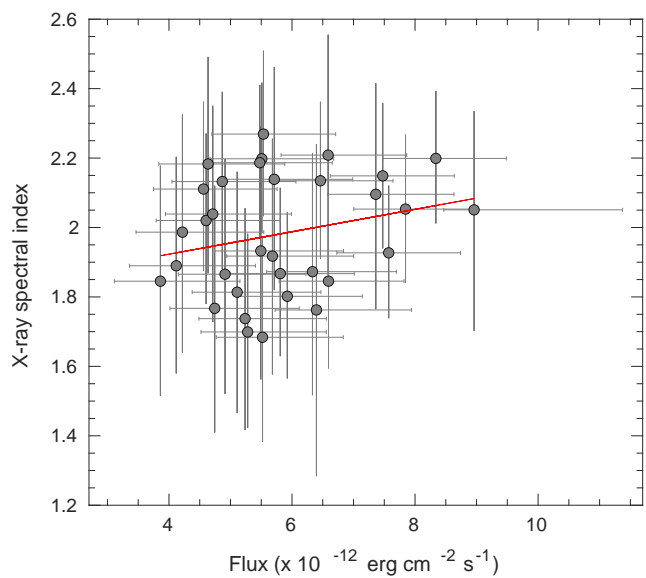}
\caption{Relation between X-ray spectral flux and slope is shown in the plot. Least-squares fit to the data, shown by the red line, gives a probability of correlation of 0.18 with a coefficient of 0.24.} 
\label{fig4}
\end{figure}

During the studied period, the spectral index of the source varies from the softest value \mbox{$\Gamma_{max} = 2.27\pm 0.24$} to the hardest value $\Gamma_{min} = 1.68\pm 0.30$ and the unabsorbed flux varied between 4.12--9.25 $\times \rm{10}^{-12}$ erg cm$^{-2}$ s$^{-1}$. Although we observed a change of source's spectral state, we did not observe any gradual change towards the outburst epoch. The X-ray spectral evolution is shown in Figure~\ref{fig3} and the relation between the absorbed X-ray flux and the spectral slope is shown in Figure~\ref{fig4} and the relation between the absorbed X-ray flux and the spectral slope is show. Even though the source is in pre-burst phase, a weak positive correlation is visible from this figure, showing the presence of a ``softer when brighter trend''. A similar trend has been reported for this source in different instances, but mostly while the source was flaring (e.g., \cite{1992AJ....104...15C, 2010MNRAS.402.2087V, 2018MNRAS.480..407K}). In the studied period, absence of an upward curvature in the spectra yields that contribution from the IC emission was not sufficient enough to dominate the synchrotron emissions coming from the jet at high energies.  

\section{Conclusions}\label{sec:summary}
\noindent

In this paper, we present the results of a temporal and spectral study of the popular blazar OJ~287 during its pre-outburst phase before the April--June 2020 outburst. In this work, we carried out a multi-wavelength (optical, UV and X-ray) variability study on a short time scale (STV), X-ray spectral analysis, and cross-correlation study using observations collected by the {\it Swift} satellite. Findings of this study are summarized below:
\begin{itemize}
\item In the pre-outburst period, the source showed short-term variability in the optical to soft X-ray (0.3--2 keV) bands at a moderate level with $F_{var}$ values ranging from 22--31\%, while the variability in the 2--10 keV hard X-ray band ($\approx$8\%) was significantly less as compared to that of others. During this period, the overall X-ray emission was significantly dominated by the soft X-ray~photons. 

\item Most of the X-ray spectra were well described by an absorbed power-law model in the 0.3--10~keV energy range. In this period, the unabsorbed X-ray flux increased by a factor of two and the spectral slope varied by $\Delta\Gamma$ = 0.6. A weak, but detectable ``softer when brighter'' spectral trend similar to that usually observed in the flaring state of the source was also observed. This could be an outcome of the same emission process that triggered the April--June 2020 outburst. Recently, \citet{2020ATel13785....1K} have reported the X-ray spectrum following the same trend during the April 2020 outburst, where the X-rays are comparable in brightness to the December 2015 X-ray flare which was triggered due to the impact of the inspiraling binary black hole system \citep{2016ApJ...819L..37V} and to the February 2017 flare which was most likely jet-powered \citep{2018MNRAS.479.1672K}.

\item Optical to soft X-ray emissions show weakly correlated variability, where the optical/UV photons are lagged behind the soft X-rays by $\approx$45--48 days. The hard X-rays did not show any correlation with the lower energy bands. These results indicate that these photons were emitted by two different electron populations; most likely, the lower energy emissions below soft X-rays were produced by the synchrotron process in the jets and the hard X-rays and above were produced by the EC process. 
\end{itemize}
\vspace{6pt} 

\authorcontributions{Original draft writing, N.K.; review and editing, A.C.G; resources, M.G. All authors have read and agreed to the published version of the manuscript.}

\funding{N.K. acknowledges funding from the Chinese Academy of Sciences President’s International Fellowship Initiative Grant No. 2020PM0029. M.G. acknowledges funding from the National Science Foundation of China Grant No. 11873073.}

\acknowledgments{We acknowledge the use of public data from the Swift data archive. We thank the referees for their useful suggestions which helped to improve the quality of the paper.}

\conflictsofinterest{The authors declare no conflict of interest.} 

\reftitle{References}

\end{document}